\documentclass[twocolumn,aps,floats,superscriptaddress,showpacs]{revtex4}
\usepackage{amsmath}
\usepackage[usenames,dvipsnames]{color}

\usepackage{epsfig}
\usepackage{graphicx}
\usepackage{dcolumn}
\usepackage{bm}
\usepackage{ dsfont }
\newcommand{\beq}{\begin{equation}}
\newcommand{\eeq}{\end{equation}}
\newcommand{\beqa}{\begin{eqnarray}}
\newcommand{\eeqa}{\end{eqnarray}}

\renewcommand{\dag}{^{\dagger}}

\def\gapp{\lower.35em\hbox{$\stackrel{\textstyle>}{\sim}$}}
\def\lapp{\lower.35em\hbox{$\stackrel{\textstyle<}{\sim}$}}

\begin{document}
\bibliographystyle{apsrev}
%


\title{Space dependent Fermi velocity in strained graphene}

\author{Fernando de Juan}
\affiliation{Department of Physics, Indiana University, Bloomington, IN 47405}
\author{Mauricio Sturla} 
\affiliation{Instituto de Ciencia de Materiales de Madrid,\\
CSIC, Cantoblanco; 28049 Madrid, Spain.}
\author{Mar\'ia A. H. Vozmediano}
\affiliation{Instituto de Ciencia de Materiales de Madrid,\\
CSIC, Cantoblanco; 28049 Madrid, Spain.}
\date{\today}
\begin{abstract}
We investigate some apparent discrepancies between two different  models for curved graphene: the one
based on tight binding and elasticity theory,
and the covariant approach based on quantum field theory in curved space. We demonstrate that
strained or corrugated samples will have a space dependent Fermi velocity  in either approach
that can affect the interpretation of local probes experiments in graphene. 
We also generalize the tight binding approach to general
inhomogeneous strain and find a gauge field proportional to the derivative of
the strain tensor that has the same form as the one obtained in the covariant approach.
\end{abstract}
%
%
%
%
 \maketitle

As it is well known the low energy electronic excitations in graphene are modeled by a massless
Dirac Hamiltonian that is able to describe  many spectroscopic and transport experiments with
surprising accuracy \cite{NGetal09}.  The Fermi velocity is the only free parameter in the model
and, as such, all the observable quantities depend critically on it. It plays the same role as the
effective mass in the usual Fermi liquid theory.  Recent experiments  \cite{LLA11,EGetal11,SPetal11}
have
been able to measure the renormalization of the Fermi velocity predicted in \cite{GGV94}  what means
that the Fermi velocity is not a constant but increases as  the energy is decreased near the Dirac
point. In this work we will show that the corrugations and strain present in most of the samples
give rise to a space-dependent Fermi velocity, a fact that can change the interpretation of some
experiments.

The presence of ripples in all the graphene samples 
and the influence of the lattice distortions on the electronic properties
of the material are two of the most interesting aspects of graphene that still remain as
open problems in the field. 
The most popular model proposed in the literature to study the issue is based on
a combination of tight binding and elasticity theory \cite{SA02,GHD08} that induces
"elastic" vector fields coupled to the electronic density (for an extensive review
on the subject see \cite{VKG10} and the references therein). This approach 
has given rise to the proposal of "strain engineering" \cite{PC09,GKG10}. An important
highlight in the field has been the observation of a reorganization of the spectrum resembling
Landau levels in strained graphene due to the effective magnetic field \cite{LBetal10} whose
existence was predicted theoretically in \cite{GKG10,GGetal10}.

An alternative model to explore the influence of corrugations on the electronic properties
of graphene is based on the formulation of quantum field theory (QFT) in curved space well
known in cosmology and gravitation \cite{BD82,W96}. 
This possibility, suggested for modeling spherical fullerenes in \cite{GGV92,GGV93}, was 
fully explored for general geometries in \cite{CV07a,CV07b,JCV07,JCV10}. The QFT approach is
rooted on the spinorial nature of the low energy electronic excitations of graphene and
on the robustness of the Fermi points under deformations of the lattice \cite{MGV07}, and
hence it should work well in the low energy scale.

Since both models are natural and predictive, one should expect that they will provide the same
results when applied to curved graphene samples with given shapes. Nevertheless there are a few 
discrepancies that are immediately apparent without entering into much detail. One is the prediction
of the QFT approach of a space--dependent Fermi velocity put forward in \cite{JCV07}. The origin
of it lies on the vector indices of the Pauli matrices that made them space dependent when going to
curved geometries. There is also a homogeneous contribution coming from volume effects. Another obvious discrepancy is related to  in--plane distortions. These give
rise in the elasticity approach to the coupling of the spinor to vector fields (named fictitious
magnetic fields in the literature) which in turn have observational consequences. Since in--plane
distortions do not induce intrinsic curvature to the sample, the analogue of the fictitious vector
fields in the QFT approach, the spin connection, is zero. A third apparent discrepancy comes from
the different symmetry of the  vector fields associated to intrinsically curved samples where the
two approaches can be used. The best example is provided by the  gaussian bump worked out in
\cite{JCV07} whose associated vector field is axially symmetric in the QFT scheme and has trigonal
symmetry in the TB-elasticity models \cite{JCetal11}.

Although the tight binding derivation of the low energy effective Dirac Hamiltonian is the most
popular probably for historical reasons \cite{W47}, it has been known since 1958 that the Dirac structure is more general and follows from the lattice symmetry and a low energy expansion  \cite{SW58}. 
New directions studying the effective low energy Hamiltonians for distorted lattices on symmetry
grounds  \cite{M07,WGS08,PL09,WZ10,Lin11} paved the way to a better understanding of the correspondence
between the lattice formulations and the QFT covariant approach. 
In the present work we partially follow this path and explore the correspondence between lattice and
continuum formulations with special emphasis on the spacial dependence of the effective Fermi
velocity. We will show that 
a space dependent Fermi velocity arises from the tight binding--elasticity approach when going beyond
the linear approximation and that the two formalisms can be compared if a metric coming from
elasticity is chosen  
in the QFT approach. 

{\bf The QFT geometric description of curved graphene and the elasticity theory.}

The QFT geometric description of curved graphene has been discussed in detail in \cite{VJC08}.  
It is based on the stability of the Fermi points of the hexagonal lattice under moderate lattice
distortions \cite{MGV07} and on the subsequent description of the low energy excitations 
around the Fermi points as massless Dirac fermions. Hence 
a natural way to incorporate the effect of the observed corrugations at low energies is to couple
the Dirac equation to the given curved background.

To make the connection between the two approaches explicit, we will work out the QFT Hamiltonian in
a curved space arising from a metric related to the strain tensor by $g_{ij} = \eta_{ij} + 2
u_{ij}$ \cite{LL70,K89}, where $\eta_{ij}$ is the flat metric (the identity matrix), and 
the strain tensor is defined as  $u_{ij} = \frac{1}{2} \left( \partial_i u_j + \partial_j u_i
+\partial_i h \partial_j
h \right)$, where $u_i$ and h are in and out of plane displacements respectively.
Since the metric $g$ is already phenomenological,  we could include in its definition a material
dependent parameter $\beta$ similar to the one obtained from the TB approach. We choose to leave its
value to 1 for the time being. 

The Dirac Hamiltonian in a curved space described by a metric $g_{ij}$ is given by
\begin{equation}
H = i \int d^2 x \sqrt{g} \bar{\psi} \gamma^a e_a^{i} (\partial_{i} +
\Omega_{i}) \psi,
\label{curveH}
\end{equation}
where $\bar{\psi}= \psi^{\dagger} \gamma^0$, $\gamma^0\gamma^i = \sigma^i$ are the Pauli matrices,
and a sum is implicit in repeated indices. The effect of the metric is encoded in the tetrads
$e_a^i$, the metric determinant $\sqrt{g}$, and the spin connection $\Omega_i$. Expanding these
objects to first order in $u_{ij}$ (see the Supplementary Material A) we obtain the
Hamiltonian 

\beqa
\label{Hcurved}
H =& i \int d^2 x \psi^{\dagger} ( \sigma_a \partial_a  + u_{ii}  \sigma_a
\partial_a- \sigma_a u_{ab}
\partial_b +\\ \nonumber
&\frac{1}{2}\sigma_a ( \partial_a u_{ii} -\partial_i u_{ia} )
)\psi.
\eeqa
The first term in eq. \eqref{Hcurved} is the usual Dirac term. The next two terms are the space
dependent Fermi velocity, and the last two are the corresponding geometric gauge field, as
interpreted in ref. \cite{JCV07}.
Note that terms of the type $i (u_{ij}
\partial_k + 1/2 \partial_k u_{ij})$ are Hermitean by partial integration: for every term in the
Fermi velocity there must be a corresponding geometric gauge field that guarantees
hermiticity. Also note the extra factor of $i$ as compared to the minimal coupling of a U(1) gauge
field, $\partial_i + i e A_i$, which is Hermitean by itself. The effective Hamiltonian around the
other Fermi point will have the same Fermi velocity but a minus sign in the gauge field couplings.

{\bf Space--dependent Fermi velocity from tight-binding.}
\label{sec_vf}

We now proceed to compute the effective Hamiltonian in the presence of strain directly from the
tight binding model. The general tight binding Hamiltonian is
\begin{equation}
H = - \sum_{\vec x,n} t_{\vec x,n} a^{\dagger}_{\vec x} b_{\vec x+\vec \delta_n}
+ cc.,
\label{TBH}
\end{equation}
where $\vec x$ runs over the position of all unit cells and the three nearest neighbour vectors are 
defined as
\begin{align}
\vec \delta_1 = a (\frac{\sqrt 3}{2}  , \frac{1}{2}) &&
\vec \delta_2 = a (-\frac{\sqrt 3}{2} , \frac{1}{2}) &&
\vec \delta_3 = a (0  , -1),
\end{align}
with $a$ the equilibrium nearest neighbour distance.

In order to make an easier contact with previous works in the literature and to illustrate the
method we will first consider the case of homogeneous
strain $t_{\vec x , n} = t_n$  and then generalize it to arbitrary inhomogeneous strain.
In the tight binding approximation, the only effect of strain is to modify the hopping integrals as
the distance between atoms changes. Even if the strain does not depend on position, the three
nearest neighbour hoppings $t_n$ may vary independently. To first order in the distance change
$\Delta u_n$, we may write
\begin{equation}
t_n = t_0( 1 - \beta \Delta u_n ) \label{tn},
\end{equation}
where $\beta = \vert\partial \log t / \partial \log a\vert$ and $t_0$ the equilibrium hopping. The
relative
distance change to first order in strain is
\begin{equation}
\Delta u_n = \frac{\delta_n^i \delta_n^j}{a^2} u_{ij} \label{un}.
\end{equation}
Notice that when considering inhomogeneous strain as done in the Supporting Information C
this equation has to be 
completed with terms proportional to the derivative of the strain tensor. 
The lowest order in this case is
\begin{equation}
\Delta u_n = \frac{\delta_n^i \delta_n^j}{a^2} u_{ij} +
\frac{\delta_n^i \delta_n^j\delta^k_n}{2 a^2} \partial_i u_{jk}.
\label{inhomogeneous}
\end{equation}
We now expand in Bloch waves
\begin{align}
a_{\vec x} = \sum_{\vec k \in BZ} e^{i \vec k \vec x} a_{\vec k}, & & b_{\vec x} = \sum_{\vec k \in
BZ} e^{i \vec k \vec x} b_{\vec k},
\end{align}
and consider momenta close to the Dirac points $\vec k = \vec K + \vec q$, with $\vec K =
(4\pi /(a3\sqrt 3),0)$ (the result for the other Dirac point can obtained changing $\vec K
\rightarrow
- \vec K$ throughout). The Hamiltonian in momentum space is
\begin{equation}
H = -\sum_{n=1}^3 t_n \left( \begin{array}{cc} 0 &  e^{-i (\vec K+ \vec q)  \cdot
\vec{\delta}_n }\\ e^{i
(\vec K+ \vec q) \cdot \vec{\delta}_n } & 0 \end{array}\right).
\end{equation}
Expanding to first order in $q$ we obtain
\begin{equation}
H \approx -\sum_{n=1}^3 t_n \left(
\begin{array}{cc} 0 &  e^{-i \vec K   \vec{\delta}_n }\\
e^{i
\vec K \vec{\delta}_n } & 0 \end{array}\right)(1+i \sigma_3   \vec q \cdot \vec \delta_n).
\end{equation}
We now use the following identity
\begin{equation}
\left( \begin{array}{cc} 0 &  e^{-i \vec K \cdot \vec{\delta}_n }\\  e^{i \vec
K \cdot \vec{\delta}_n } & 0 \end{array}\right) =  i \frac{\vec{\sigma} \cdot
\vec{\delta}_n}{a} \sigma_3,
\label{identity}
\end{equation}
where $\vec \sigma = (\sigma_x ,\sigma_y)$ are the two Pauli matrices. Using eqs.
(\ref{tn}) and (\ref{un}) for $t_n$ the Hamiltonian is 
\begin{equation}
H \approx -\sum_{n=1}^3 t_0 \left(1+\frac{\beta}{a^2} \vec \delta_n u \vec \delta_n \right)
\left(\frac{i}{a}\sigma_3
\vec{\sigma} \cdot \vec{\delta}_n\right)\left(1+i \sigma_3   \vec q \cdot \vec \delta_n\right).
\end{equation}
We can now collect the different terms of this expression with the use of the identities given in
the Supplementary Material B.
Labeling the various terms by their order in the expansion in
$q,u$ we get
\begin{equation}
H = H_q + H_u + H_{q,u},
\end{equation}
with
\begin{align}
H_q &=  v_0 \sigma_i q_i,  \\
H_u &= \frac{v_0}{2a} \beta \sigma_i K_{ijk} \epsilon_{kl} u_{jl}, \\
H_{u,q} &= \frac{v_0}{4}  \beta \left[ 2 \sigma^i q_j u_{ij} + \sigma^i q_i u_{jj} \right],
\label{new}
\end{align}
where we have defined $v_0 = 3t_0 a/2$ and $K_{ijk}$  is the invariant $C_3$ tensor given in
\eqref{K} \cite{VKG10}.
It is easy to see that, if needed, the expansion of the low
energy hamiltonian in powers of $q_{i}$
and $u_{ij}$ can be done to any order. The first two contributions are well known: 
$H_q$ is the usual Dirac Hamiltonian, and  $H_u$ is the standard strain induced
gauge coupling  \cite{VKG10}:
$H_u = v_0 \sigma_i A_i$ 
with components
\beq
A_x = \frac{\beta }{2 a} (u_{xx}-u_{yy})\;\;\;,\;\;\;
A_y = \frac{\beta }{2 a} (-2u_{xy}).
\label{standard}
\eeq
The new contribution $H_{u,q}$ is the main result of this work. This result is consistent with the
one obtained from the symmetry analysis in \cite{WZ10,Lin11}, and fixes the coefficients of the
symmetry allowed terms in terms of a microscopic model. We will see its full significance in what
follows.

{\bf Inhomogeneous strain.}
Treating inhomogeneous strain is a delicate issue due to the lack of translation invariance of the
system. The usual procedure \cite{Lin11} consists in taking the homogeneous Hamiltonian in k space
and go to real space by the replacement rule 
\begin{equation}
u_{ij} q_k \rightarrow i(u_{ij} \partial_k + \frac{1}{2} \partial_k u_{ij}),
\label{rule}
\end{equation}
which guarantees Hermiticity. Using this rule, the  term  \eqref{new} becomes
\begin{equation}
H_{u,q} = i \frac{v_0}{4} \sigma_i \left( 2 u_{ij} \partial_j + u_{jj} \partial_i +
\partial_j u_{ij}+ \frac{1}{2}\partial_i u_{jj}\right),
\end{equation}
what allows to write the total Hamiltonian as
\begin{equation}
H = i  v_{ij}(r) \sigma_i \partial_j + i v_0 \sigma_i \Gamma_i  + v_0 \sigma_i A_i,
\end{equation}
The field $A_i$ is the one obtained in the standard approach \eqref{standard}. We also get
the tensorial and space dependent Fermi velocity
\begin{equation}
v_{ij} = v_0 \left[\eta_{ij} + {\frac{ \beta}{4}}(2 u_{ij} + \eta_{ij} u_{kk})\right],
\label{vfermi}
\end{equation}
obtained in \eqref{new}. The new term is 
a ``geometric" gauge field given by
\begin{equation}
\Gamma_i = \frac{\beta}{4} \left(\partial_j u_{ij}+ \frac{1}{2}\partial_i u_{jj}\right).
\label{Aprima}
\end{equation}
which was obtained in the covariant approach \eqref{Hcurved}. Being proportional to the
derivative of the strain tensor it only appears in the case of inhomogeneous strain.
All these terms are also found in the symmetry analysis \cite{WZ10,Lin11}. 

As discussed in \cite{Z94}, the former procedure  can miss some terms. We have checked that this
is the complete result to this order in derivatives by directly performing the
Fourier transform of the real space Hamiltonian in the first stage of the tight binding for
the general case of inhomogeneous strain. The
details can be found in the Supporting Information C.

{\bf Discussion and future.}

The purpose of this work was to discuss the equivalence of the tight binding and the QFT approaches
in the description of curved or strained graphene with special emphasis on the space dependent Fermi
velocity. The main result is that the two approaches give rise to the same type of terms for the
spacial dependent Fermi velocity although with  different numerical values of the coefficients. 
The symmetry approach gives rise to the same type of terms with independent - and undetermined - 
coefficients whose value has to be fixed by the model.  We have also derived the "geometric" gauge field
in the tight binding approach for the case of inhomogeneous strain.

Referring to the QFT approach, we have seen that the definition of  the metric of the curved space
in terms of the strain tensor allows to get physical effects from the covariant formalism even in
the case of having only in-plane distortions. 

An interesting related issue concerns whether new terms can appear in the effective tight binding
Hamiltonian independently of the change of the hoppings, produced just by changes in the relative
positions of the atoms. If one assumes $\beta=0$, the discrete TB Hamiltonian in that case is just 
\begin{equation}
H = - \sum_{<ij>} t_0 a^{\dagger}_{i} b_{j} + cc.,
\end{equation}
with $i,j$ running through nearest neighbour atoms. It is easy to see that moving the atoms form
their positions does not change the Hamiltonian at all. $i,j$ are just labels numbering the atoms,
and need not refer to physical position in any sense. 
Hence the energies and eigenfunctions  of the system are not modified by this strain, even if
inhomogeneous (for $\beta$=0). However, this does not imply that there will not be observable
consequences. The fact that the atoms are in different positions does change the way in which
external position--dependent probes see
the system. In short, when the atoms are displaced, the discrete label $i$ maps to a physical
position in the strained frame.
If we describe the physics in the lab frame so as to be able to couple an external field, vectors in
the strained frame have to be rotated to the lab frame. This gives rise to  a $\beta$ independent
contribution of the type obtained recently in \cite{KPetal12} and will be discussed in detail
elsewhere.  

Although the  space dependent Fermi velocity is obtained at higher order in a tight binding
expansion, its presence has important physical consequences and it can not be obviated. The
important issue in this discussion is  the influence of the various factors entering the effective
Hamiltonian on the observable physical quantities. In particular in \cite{JCV07} it was shown that
in the QFT approach the local density of states (LDOS) is not affected by the vector fields to first
order in perturbation theory. The oscillations in the LDOS obtained in that work came  from the
combination of variable Fermi velocity and volume effects. Similarly it has been shown in
\cite{FCO11} that long range correlated vector fields do not alter the minimal conductivity of
graphene that is in turn severely changed by long range correlated disorder in the form of a random
distribution of  Fermi velocity \cite{CV09}. We note also that while this paper was completed we
learned on an experiment that points to  the observation of a 5-10\% spatial fluctuation of the
Fermi velocity in samples on SiO$_2$ \cite{LLA11}.

As we have seen, the standard QFT approach is rooted in considering the spinor describing the low
energy electronic excitations of graphene as a covariant spinor under a geometric point of view.
Since the time coordinate remains ``flat", the Lorentz symmetry is reduced to translation and
rotations. This approach has been pushed forward in \cite{IL11} considering graphene as a QFT in
curved {\it spacetime}  what gives rise to very interesting consequences as the possibility to
observe a Hawking-Unruh temperature in the curved graphene samples.

We thank J. M. B. Lopes dos Santos for interesting discussions in the early stages of this work. We
also acknowledge conversations with E. V. Castro, A. Cortijo, A. Grushin, F. Guinea and H. Ochoa. 
We particularly thank J. L. Ma\~nes for explaining the symmetry approach to us and for
pointing out a mistake in a previous version of the manuscript.
This research was supported by the Spanish MECD grants FIS2008-00124, PIB2010BZ-00512 and by NSF
grant DRM-1005035.

\bibliography{PRL2}
\section{Supporting information}
\subsection{The QFT geometric description of curved graphene and the elasticity theory.---}
\label{covariant}

In this section we will work out the various geometric objects entering in the curved space
Hamiltonian in eq. (1) of the main text.

The volume element will be given by $\sqrt{g} \equiv \sqrt{det g_{ij}} = 1 + u_{ii}$. 
The next geometric object needed is the tetrad, which satisfies
\begin{equation}
g_{ij} = e_i^a e_j^b\delta_{ab} = e_i^a e_j^a,
\end{equation}
The easiest choice is 
\begin{align}
e_{ai} = \delta_{ai} + \delta_a^ju_{ij}, \\
e_{a}^i = \delta_{a}^i -\delta_{aj}u^{ij},
\end{align}
Next we need the affine connection
\begin{equation}
\Gamma_{ij}^k=\frac{1}{2}g^{lk}\{\frac{\partial
g_{jl}}{\partial x^i}+\frac{\partial g_{il}}{\partial
x^j}-\frac{\partial g_{ij}}{\partial x^l}\}, 
\label{christoffel}
\end{equation}
which to first order is 
\begin{equation}
\Gamma_{ij}^k=\delta^{lk}\{\partial_i
u_{jl}+\partial_j
u_{il}-\partial_l
u_{ij}\},
\end{equation}
Next we need the spin connection, which is obtained from 
\begin{equation}
\Omega_i(x)=\frac{1}{4}\gamma_a\gamma_b
e^a_j g^{jk} \nabla_i e^b_k = -\frac{1}{4}\sigma_a\sigma_b
e^a_j g^{jk} \nabla_i e^b_k,
\end{equation}
with
\begin{equation}
\nabla_i e^a_j=\partial_i
e_j^a-\Gamma_{ij}^k e_k^a.
\end{equation}
To first order the spin connection is
\begin{equation}
\Omega_i = -\frac{1}{4} \sigma_a \sigma_b ( \partial_a u_{ib} - \partial_b u_{ia}).
\end{equation}
With aid of the identity
\begin{equation}
\sigma^a\sigma^b \sigma^c  = \sigma^a \delta^{bc} + \sigma^c \delta^{ba} - \sigma^b\delta^{ac} +
i\epsilon^{abc},
\end{equation}
we can compute
\begin{equation}
\sigma^a e_a^i \sigma_i =- \frac{1}{2}\sigma_a (\partial_i u_{ia} - \partial_a u_{ii} ).
\end{equation}
The Dirac Hamiltonian to first order in $u$ finally reads
\beqa
H =& i \int d^2 x \psi^{\dagger} ( \sigma_a \partial_a  + u_{ii}  \sigma_a
\partial_a- \sigma_a u_{ab}
\partial_b +\\ \nonumber
&\frac{1}{2}\sigma_a ( \partial_a u_{ii} -\partial_i u_{ia} )
)\psi.
\eeqa

\subsection{Some tensor identities}
\label{sums}
The following identities are used in the tight binding derivation:
\beq
i \sigma_i
\sigma_3 = \epsilon_{ij} \sigma_j
\eeq 
\begin{align}
\frac{1}{a^2}\sum_{n=1}^3 \delta_n^i \delta_n^j &= \frac{3}{2} \eta^{ij}, \\
\frac{1}{a^3}\sum_{n=1}^3 \delta_n^i \delta_n^j \delta_n^k &= -\frac{3}{4} K^{ijk}, \\
\frac{1}{a^4}\sum_{n=1}^3 \delta_n^i \delta_n^j \delta_n^k \delta_n^l &= \frac{3}{8}
(\eta^{ij}\eta^{kl} +
\eta^{ik}\eta^{jl} + \eta^{il}\eta^{jk}).
\label{formulas}
\end{align}
The object $K^{ijk}$ in the second equation in \eqref{formulas} is an invariant tensor under the
discrete $C_3$ rotations of the lattice
given by 
\begin{equation}
K^{abc}=\sum_{n=1}^3 d_i^a d_i^b d_i^c = d_1^a d_1^b d_1^c + d_2^a d_2^b d_2^c + d_3^a d_3^b d_3^c,
\label{K}
\end{equation}
with $d_i$ the nearest neighbour vectors. It is easy to see that  the only non--zero
components of it are 
\begin{equation}
K^{111} = -K^{122} = -K^{212} = -K^{221} = 1.
\end{equation}

\subsection{ Inhomogeneous strain with explicit Fourier transform.}
\label{fourier}
We start with the standard tight binding Hamiltonian
\begin{equation}
H = - \sum_{\vec x,n} t_{\vec x,n} a^{\dagger}_{\vec x} b_{\vec x+\vec \delta_n}
+ hc.,
\end{equation}
where $\vec x$ runs over the position of all unit cells and the three nearest neighbour vectors are 
defined as
\begin{align}
\vec \delta_1 = a (\frac{\sqrt 3}{2}  , \frac{1}{2}) &&
\vec \delta_2 = a (-\frac{\sqrt 3}{2} , \frac{1}{2}) &&
\vec \delta_3 = a (0  , -1),
\end{align}
with $a$ the equilibrium nearest neighbour distance. 
We now replace $a_x$ and $b_x$ by their fourier expansions and get
\beqa
&H= - \sum_{\vec x,n} \sum_{k,k'} t_{\vec x,n} \\ \nonumber
&\left( e^{-i \vec k\vec x}a^{\dagger}_k e^{i \vec
k'(\vec x  + \vec \delta_n)}b_{k'} +e^{-i \vec
k'(\vec x  + \vec \delta_n)} b^{\dagger}_{k'}e^{i \vec k\vec x} a_{k}\right). 
\eeqa
Using the expressions $u_k=\frac{1}{N} \sum_x e^{-i k x } u_x$, and  
$\sum_x e^{ikx} = N \delta_{k,0}$,
H becomes
\beqa
&H=- \sum_{\vec x,n} \sum_{k,k'} t_{\vec x,n} \\ \nonumber
& \left( e^{-i \vec k\vec x}a^{\dagger}_k e^{i \vec
k'(\vec x  + \vec \delta_n)}b_{k'} + 
e^{-i \vec k'(\vec x  + \vec \delta_n)} b^{\dagger}_{k'}e^{i \vec k\vec x} a_{k}\right).
\eeqa
To write this as a matrix equation we have to relabel $k\leftrightarrow k'$ in the first term, so
that we get
\begin{equation}
H =- N \sum_{k,k',n}  \left(a^{\dagger}_{k'}, b^{\dagger}_{k'}\right)t_{k'-k,n}
\left(\begin{array}{cc}
       0  & e^{i \vec k \vec \delta_n} \\
e^{- i \vec k' \vec \delta_n} & 0 
      \end{array}
\right)
\left(\begin{array}{c} 
       a_k \\ b_k 
      \end{array}
\right).
\end{equation}
Now we proceed as usual, expanding $\vec k = \vec K + \vec q$ and $
\vec k' = \vec K + \vec q'$. We get 
\begin{equation}
H =- N \sum_{q,q',n}  \left(a^{\dagger}_{q'}, b^{\dagger}_{q'}\right)t_{q'-q,n}
\left(\begin{array}{cc}
       0  & e^{i  (\vec K +\vec q)  \vec \delta_n} \\
e^{- i  (\vec K+\vec q') \vec \delta_n} & 0 
      \end{array}
\right)
\left(\begin{array}{c} a_q \\ b_q \end{array}\right).
\end{equation}

\begin{widetext}
To expand in $q,q'$ we rewrite this as
\begin{equation}
H =- N \sum_{q,q',n}  \left(a^{\dagger}_{q'}, b^{\dagger}_{q'}\right)t_{q'-q,n}
\left(\begin{array}{cc}
       0  & e^{i  (\vec K +\frac{1}{2}(\vec q+\vec q') + \frac{1}{2}(\vec q-\vec q'))  \vec
\delta_n} \\
e^{- i  (\vec K+\frac{1}{2}(\vec q+\vec q') - \frac{1}{2}(\vec q-\vec q')) \vec \delta_n} & 0 
      \end{array}
\right)
\left(\begin{array}{c} a_q \\ b_q \end{array}\right),
\end{equation}
and using the identity  
\begin{equation}
\left( \begin{array}{cc} 0 &  e^{-i \vec K \cdot \vec{\delta}_n }\\  e^{i \vec
K \cdot \vec{\delta}_n } & 0 \end{array}\right) =  i \frac{\vec{\sigma} \cdot
\vec{\delta}_n}{a} \sigma_3,
\label{identity}
\end{equation}
we get
\begin{equation}
H =- \frac{N}{a} \sum_{q,q',n}  c^+_{q'} t_{q'-q,n}
 (i\sigma_3 \vec \sigma \vec \delta_n) \left(1-i\sigma_3 (\frac{\vec q + \vec q'}{2}) \vec \delta_n
 + i (\frac{\vec q - \vec q'}{2})\vec \delta_n\right) c_q ,
\end{equation}
where $c_k  = (a_k,b_k)$. The modified hopping can now be expressed as the Fourier
transform of eqs. (5), (7) of the main text as:

\beq
 t_{q'-q,n}= t_0 \left( \delta_{q',q} - \beta u_{q'-q,ij}\frac{\delta_j^n \delta_i^n}{a^2} - 
i\,\beta (q'-q)_k u_{q'-q,ij}\frac{\delta_n^i \delta_n^j\delta^k_n}{2 a^2} \right).
\eeq
It is easy to see that the free Dirac term (linear in q and zeroth order in u)
and the standard TB  gauge field $A$ (linear in u and k=K) are recovered from this
expression. The new gauge field comes from the crossed term linear in $u$ and $q$:

\beqa
H_{u,q}&=& \frac{3 t_0 a \beta N }{16 }\,\sum_{q,q'} c\dag_{q'} \, u_{q'-q,ij} i \sigma_3 \sigma_k
(-i\sigma_3(q_l+q'_l)+i(q_l-q'_l))(\eta^{ij}\eta^{kl}+\eta^{ik}\eta^{jl}+\eta^{il}\eta^{jk}) c_q
\nonumber\\
&+&\frac{3 t_0 a \beta N }{16  }\,\sum_{q,q'} c\dag_{q'}\; u_{q'-q,ij}\;  \sigma_3\,\sigma_k (q_l-q'_l)
\,(\eta^{li}\eta^{kj} +\eta^{lj}\eta^{ki}+\eta^{lk}\eta^{ij})  c_{q}.
\eeqa
\end{widetext}

The terms with $q-q'$ are equal and with opposite signs, so they cancel each other. The fourier transform back to real space is done with the prescription 
\beq
\frac{(q+q')_i u(q'-q)_{jk}}{2}   \rightarrow \;i (u_{jk} \partial_i + \frac{1}{2}\partial_i
u_{jk})
\eeq
obtained from the definition of the Fourier transform. This allows to write the crossed term in the
form
\beq
H_{u,q}=i \frac{v_0 \beta}{4}(2u_{ij}+\eta_{ij} u_{kk})\sigma_i \partial_j +  iv_0 \sigma_i
\Gamma_i,
\eeq
with
\begin{equation}
\Gamma_i = \frac{\beta}{4} \left(\partial_j u_{ij}+ \frac{1}{2}\partial_i u_{jj}\right).
\end{equation}

\end{document}